\documentclass[]{aipproc}

\layoutstyle{6x9}

\SetInternalRegister\hbadness{8000} 

\begin{document}

\title{General Relativistic Simulations of Stellar Core Collapse
and Postbounce Evolution with Boltzmann Neutrino Transport}


\author{M. Liebend\"orfer}{
  address={Department of Physics and Astronomy, University of Tennessee, Knoxville, TN-37996 \\
and Physics Division, Oak Ridge National Laboratory, Oak Ridge, TN-37831},
  email={liebend@utk.edu}
}

\author{O.~E.~B. Messer}{
  address={Department of Physics and Astronomy, University of Tennessee, Knoxville, TN-37996 \\
and Physics Division, Oak Ridge National Laboratory, Oak Ridge, TN-37831},
  email={bmesser@utk.edu},
}

\author{A. Mezzacappa}{
  address={Physics Division, Oak Ridge National Laboratory, Oak Ridge, TN-37831},
  email={mezz@mail.phy.ornl.gov},
}

\author{W.~R. Hix}{
  address={Department of Physics and Astronomy, University of Tennessee, Knoxville, TN-37996 \\
and Physics Division, Oak Ridge National Laboratory, Oak Ridge, TN-37831},
  email={raph@mail.phy.ornl.gov},
}

\copyrightholder{American Institute of Physics}
\copyrightyear  {2001}

\begin{abstract}
We present self-consistent general relativistic
simulations of stellar core collapse, bounce,
and postbounce evolution for 13, 15, and 20 solar
mass progenitors in spherical symmetry. Our
simulations implement three-flavor Boltzmann neutrino
transport and standard nuclear physics. The
results are compared to our corresponding simulations
with Newtonian hydrodynamics and O(v/c) Boltzmann
transport.
\end{abstract}

\maketitle

\section{Introduction}

A supernova explosion is a dramatic event that includes such
a rich diversity of physics that, with current computer hardware,
a self-consistent model based on numerical simulations
can not possibly include all of them at once.
After stellar core collapse, a compact object at the center
of the event is formed, requiring a description in general
relativity. Neutrinos radiating from this central object are
strongly coupled to the matter at high densities before streaming out
at lower densities. Multi-frequency radiation hydrodynamics must
be used to quantify the energy that these
neutrinos deposit in the material behind the shock.
Moreover, evidence suggests that this heating drives
convection behind the shock,
and significant rotation and strong magnetic fields might
also be present. Observations of neutron star kicks, mixing of species,
inhomogeneous ejecta, and polarization of spectra
support the presence of asymmetries in supernova explosions
(e.g. Tueller et al. \cite{Tueller_et_al_91},
Strom et al. \cite{Strom_et_al_95},
Galama et al. \cite{Galama_et_al_98},
Leonard et al. \cite{Leonard_et_al_00},
and references therein).
Motivated by such observations, various
multi-dimensional explosion mechanisms have been explored
(Herant et al.\cite{Herant_Benz_Colgate_92},
Miller et al.\cite{Miller_Wilson_Mayle_93},
Herant et al.\cite{Herant_et_al_94},
Burrows et al.\cite{Burrows_Hayes_Fryxell_95},
Janka and M\"uller \cite{Janka_Mueller_96},
Mezzacappa et al.\cite{Mezzacappa_et_al_98b},
Fryer \cite{Fryer_99},
Fryer and Heger \cite{Fryer_Heger_00}) and jet-based
explosion scenarios have recently received new momentum
(H\"oflich et al. \cite{Hoeflich_Wheeler_Wang_99},
Khokhlov et al. \cite{Khokhlov_et_al_99},
MacFadyen and Woosley \cite{MacFadyen_Woosley_99},
Wheeler et al. \cite{Wheeler_et_al_00}). Many
exciting aspects of jets are
discussed in these proceedings.
However, for the time being, one has to single out a
subset of the known physics and to investigate the role
each part plays in a restricted simulation. It is
natural to start with
ingredients that have long been believed to be essential
for the explosion and to add modifiers in a systematic
way until the observables can be reproduced
(see e.g. Mezzacappa et al. \cite{Mezzacappa_et_al_01}).
Spherically symmetric supernova
modeling has a long tradition
and is nearing a definitive point in the sense that
high resolution hydrodynamics, general relativity,
complete Boltzmann neutrino transport, and reasonable nuclear
and weak interaction physics are being combined to dispel
remaining uncertainties.

Our simulations are based on the three-flavor Boltzmann solver
of Mezzacappa and Bruenn \cite{Mezzacappa_Bruenn_93a,
Mezzacappa_Bruenn_93b,Mezzacappa_Bruenn_93c} that was consistently
coupled to hydrodynamics on an adaptive grid and extended to
general relativistic flows (Liebend\"orfer \cite{Liebendoerfer_00}).
We performed relativistic simulations for progenitors with different
masses and compare the results to the counterpart simulations in
Newtonian gravity with O(v/c) Boltzmann transport (Messer \cite{Messer_00}).
While the quantitative results are sensitive to the inclusion of
all neutrino flavors and general relativity
(Bruenn et al. \cite{Bruenn_DeNisco_Mezzacappa_01}),
we find the same qualitative outcome as did Rampp and Janka
\cite{Rampp_Janka_00}
in their independent simulations with one-flavor O(v/c) Boltzmann
transport: Given current input physics, there are no explosions
in spherical symmetry without invoking multidimensional effects
(Mezzacappa et al. \cite{Mezzacappa_et_al_01},
Liebend\"orfer et al. \cite{Liebendoerfer_et_al_01}).
This is consistent with the results of Wilson and Mayle
\cite{Wilson_Mayle_93}, who found explosions only by including
neutron finger convection.

\section{Core Collapse and Neutrino Burst}

The simulations discussed herein are initiated from progenitors
with main sequence masses of $13$ M$_{\odot}$, $20$ M$_{\odot}$
(Nomoto and Hashimoto \cite{Nomoto_Hashimoto_88}), and $15$ M$_{\odot}$
(Woosley and Weaver \cite{Woosley_Weaver_95}). We use in our models
the equation of state of Lattimer and Swesty \cite{Lattimer_Swesty_91}
and ``standard'' weak interactions (e.g. Bruenn \cite{Bruenn_85}).
The first phase in the simulations leads through core collapse
to bounce and shock formation.
When the shock passes the neutrinospheres approximatively $4$ ms
after bounce, an energetic neutrino burst
is released from the hot shocked material, rendering it
``neutrino-visible'' to the outside world.
The corresponding deleptonization behind the shock is dramatic.
The energy carried off with the neutrino burst
enervates the shock in both the NR and the GR cases.
A pure accretion shock continues to propagate outwards
as infalling material is dissociated and layered on
the hot mantle. This stage is the definitive
end of a ``prompt'', i.e. purely hydrodynamic, explosion.

\begin{figure}
  \includegraphics{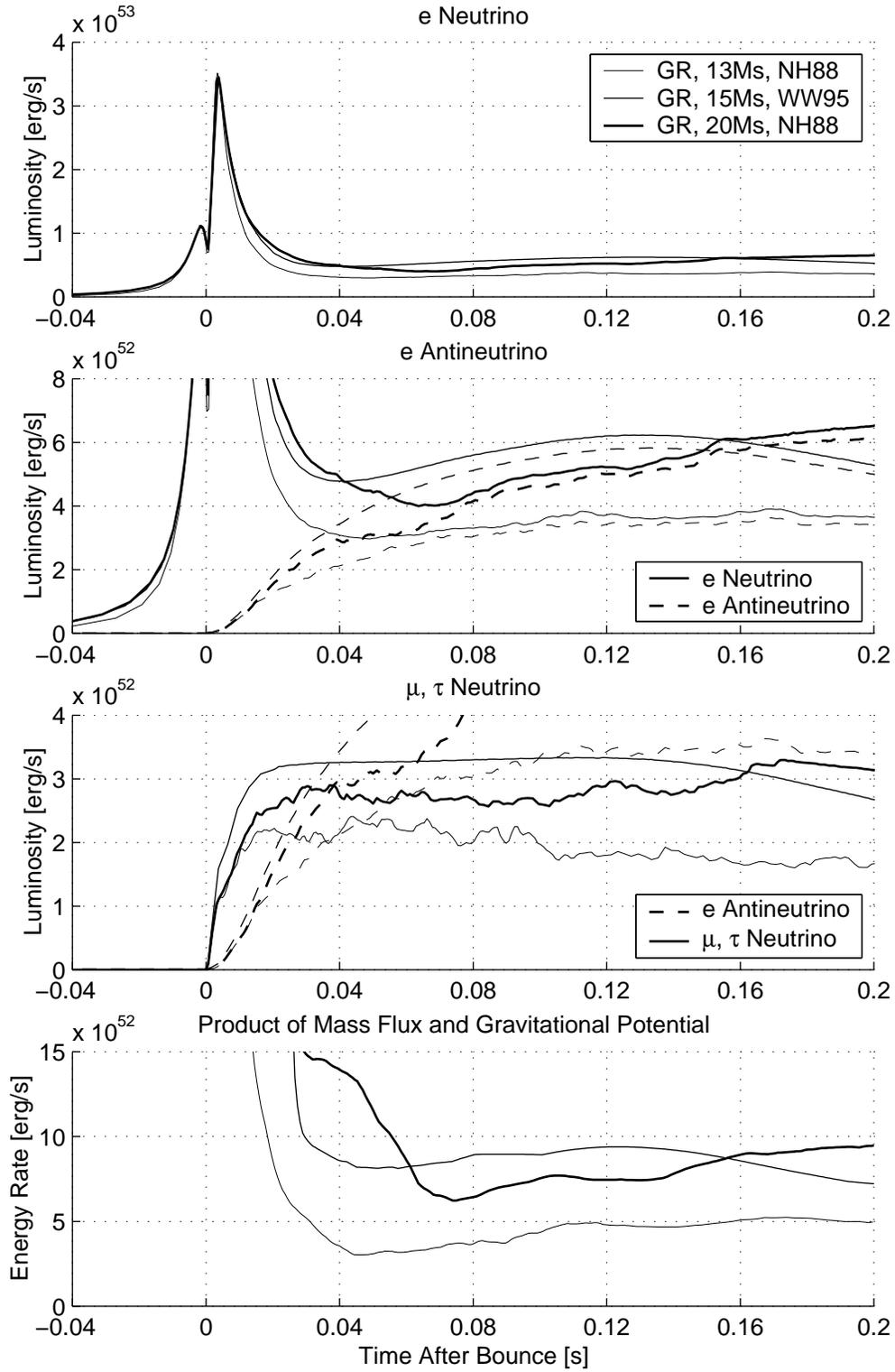}
  \caption{Luminosities for all flavors in the GR simulations. The
           last graph shows an energy rate calculated by the product
           of the gravitational potential at the neutrinosphere and
           the mass flux crossing it.}
  \label{luminzoo.ps}
\end{figure}

The neutrino luminosities from the general relativistic
simulations are shown in Fig. (\ref{luminzoo.ps}).
The electron neutrino luminosity slowly rises
during collapse and decreases as the core reaches
maximal density. It remains suppressed for the
$\sim 4$ ms the shock needs to propagate to the
electron neutrinosphere. The most prominent feature is
the electron neutrino burst, reaching $3.5\times 10^{53}$
erg/s at shock breakout, and declining afterwards.

\begin{figure}
  \includegraphics{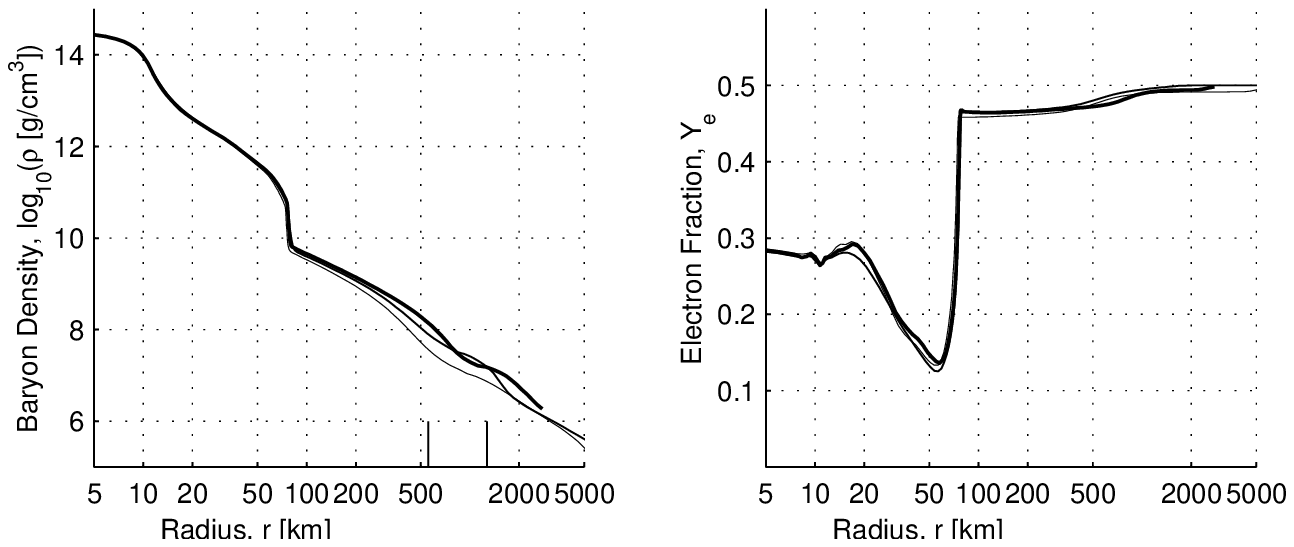}
  \caption{Similar core profiles in the GR cases at $10$ ms after bounce.}
  \label{similar.ps}
\end{figure}

The initial similarities in density and temperature 
in the inner parts of the three cores are responsible for the
similar evolution of each core with respect to collapse,
bounce, and the signature of the electron neutrino burst
(Messer \cite{Messer_00}). This is shown in Figs.
(\ref{luminzoo.ps}) and (\ref{similar.ps}).
Any initial difference in the electron fraction, $Y_e$,
in the inner part of each core cannot survive in regions that reach
sufficient densities for neutrino trapping and equilibration to occur.
If a mass element reaches neutrino trapping density,
the total lepton fraction, $Y_{l}$, becomes the
variable determining the state of the mass element.
At equilibrium, the electron capture and neutrino absorption rates are
related to one another through detailed balance, and the final $Y_{e}$
becomes a function of the local temperature and density through the
electron chemical potential. Since the inner temperature and density
profiles differ very little between the models,
each settles to essentially the same equilibrium  $Y_{e}$ deep in
the core. This determines the size of the homologous inner core and,
consequently, the position of shock formation. Since this position
sets the amount of mass that has to be dissociated when the shock
ploughs to the neutrinospheres, it strongly influences the shock
energy. Thus, a strikingly similar shock initiation is found for
the three progenitors. However, the initially dissociated mass differs
largely between the GR and NR cases because of the smaller enclosed
mass at shock formation
due to the GR effects in the gravitational potential
(GR: $0.53$ M$_{\odot}$, NR: $0.65$ M$_{\odot}$).
The luminosities in the GR and NR cases are very similar for
the electron neutrino burst. Afterwards, the GR luminosities
are generally $10\%-20\%$ larger than the NR luminosities.


\section{Heating and Accretion}

In the standard picture, the ensuing evolution is driven
by electron neutrino heating. Electron flavor
neutrinos diffuse out of the cold unshocked core
and are created in the accreting and compactifying matter around the
neutrinospheres in a hot shocked mantle.
If the heating is not sufficient to revive the shock
within the first $\sim 200$ ms after bounce (as is the
case in our simulations), the material in the heating
region is drained from
below (Janka \cite{Janka_00}), and the conditions for heating
deteriorate because of the shortened time the infalling
matter spends in the heating region. We observe this process
earlier in the general relativistic
simulations than in the
nonrelativistic simulations.

\begin{figure}
  \includegraphics{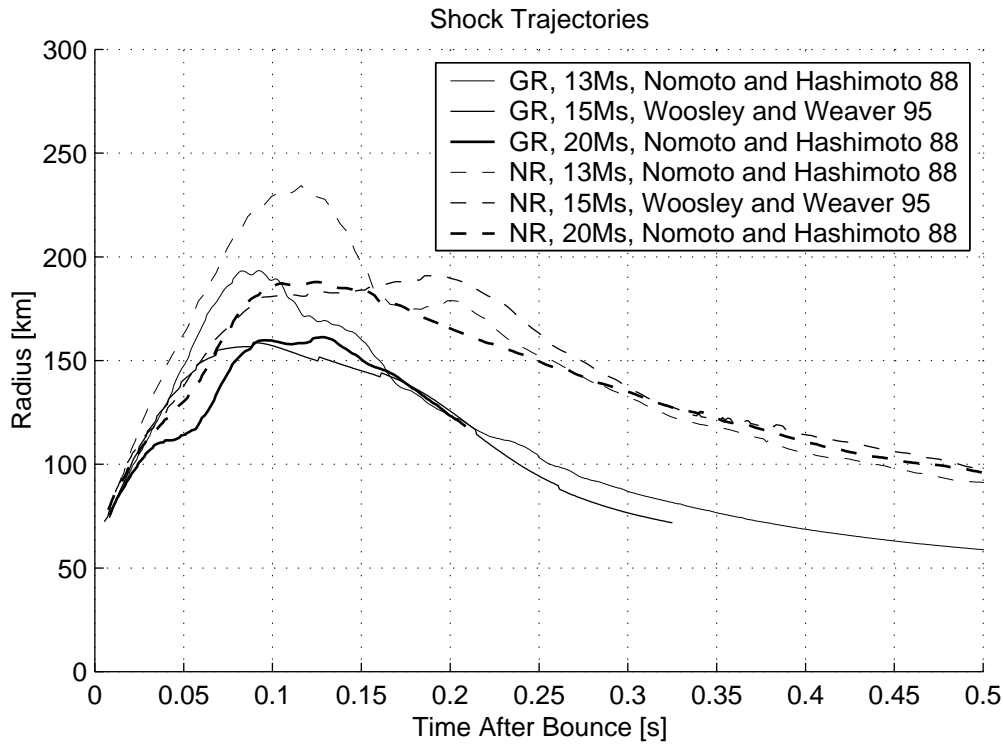}
  \caption{Shock trajectories for the GR and NR cases.}
  \label{shockzoo.ps}
\end{figure}

The shock trajectories for the different progenitors are
shown in Fig. (\ref{shockzoo.ps}).
The main difference between the GR and NR simulations stems
from the difference in the size of the proto-neutron
star, which is caused by the nonlinear GR effects at very high
densities in the center of the
star. At radii of order $100$ km and larger,
the gravitational potential becomes comparable
in the GR and NR cases. However, large differences arise if the
steep gravitational well is probed at different
{\em positions} in the GR and NR simulations, as
happens with accretion down to the neutrinospheres.
The deeper neutrinospheres in the more compact GR case encounter
material that has traversed a larger potential difference,
producing more energetic accretion luminosities.
After the shock has stalled, the surrounding layers
(cooling/heating region, shock radius) adjust to a
smaller radius and settle to a stationary equilibrium
in the spirit of Burrows and Goshy \cite{Burrows_Goshy_93}.
In the GR simulation, this occurs at a smaller radius,
deeper in the gravitational well, with higher infall
velocities, leading to higher accretion luminosities with higher
rms energies and higher heating rates (see also Bruenn et al.
\cite{Bruenn_DeNisco_Mezzacappa_01}).

This accretion-determined picture obtains additional support
from the following observation: In the last graph in
Fig. (\ref{luminzoo.ps})
we provide the mass flux at the electron neutrinosphere
multiplied by the local gravitational potential.
We can compare this energy rate to the luminosities of the electron
flavor neutrinos that are shown in the second graph in Fig. (\ref{luminzoo.ps}).
The striking similarity suggests that the luminosity is indeed
mainly determined by the gravitational potential at the boundary
of the proto-neutron star and the mass accretion rate. Moreover,
we can directly relate the different accretion rates to differences in
the density profiles of the initial progenitors.
The energy accumulation rate at the neutrinospheres for the
$15$ M$_{\odot}$ model exceeds that of the $20$ M$_{\odot}$ model
between $60$ ms and $160$ ms after bounce. We have marked in Fig.
(\ref{similar.ps}) the region that crosses the neutrinospheres
during this time. The density of the $15$ M$_{\odot}$
progenitor exceeds the density of the $20$ M$_{\odot}$ progenitor
in exactly this region.
Thus, the time dependent electron flavor luminosities
reflect the position of the neutrinospheres and are directly
modulated by the variations in the spatial density profiles of
the progenitors.


\begin{theacknowledgments}
M.L. is supported by the National Science Foundation under contract 
AST-9877130 and, formerly, was supported by the Swiss National 
Science Foundation under contract 20-53798.98. 
O.E.M. is supported by funds from the Joint Institute for Heavy Ion 
Research and a Dept. of Energy PECASE award.
A.M. is supported at the Oak Ridge National Laboratory, managed by
UT-Battelle, LLC, for the U.S. Department of Energy under contract
DE-AC05-00OR22725. 
W.R.H. is supported by NASA under contract NAG5-8405 and by funds 
from the Joint Institute for Heavy Ion Research.
Our simulations were carried out on the ORNL Physics Division Cray J90 
and the National Energy Research Supercomputer Center Cray J90. 

\end{theacknowledgments}


\end{document}